\documentclass[12pt]{iopart}

\usepackage{graphicx}

\begin{document}

\title{A unidirectional invisible $\mathcal{PT}$-
symmetric complex crystal with arbitrary thickness}

\author{Stefano Longhi}

\address{Dipartimento di Fisica and Istituto di Fotonica e Nanotecnologie del Consiglio Nazionale delle Ricerche,
Politecnico di Milano, Piazza L. da Vinci 32, I-20133 Milano, Italy}
\ead{longhi@fisi.polimi.it}
\begin{abstract}
We introduce a new class of $\mathcal{PT}$-symmetric complex crystals which 
are almost transparent and one-way reflectionless over a broad frequency range around the Bragg frequency, i.e. unidirectionally invisible, regardless of the thickness $L$ of the crystal.  The $\mathcal{PT}$-symmetric complex crystal is synthesized by a supersymmetric transformation 
of an Hermitian square well potential, and exact analytical expressions of transmission and reflection coefficients are given. As $L$ is increased, the transmittance and reflectance from one side remain close to one and zero, respectively, whereas the reflectance from the other side secularly grows like $ \sim L^2$ owing to unidirectional Bragg scattering. This is a distinctive feature as compared to the previously studied case of the complex sinusoidal $\mathcal{PT}$-symmetric potential $V(x)=V_0 \exp(-2i k_ox)$ at the symmetry breaking point, where transparency breaks down as $L \rightarrow \infty$.
\end{abstract}


\maketitle

\section{Introduction}
Over the past two decades a great and increasing interest has been devoted to study the transport and scattering properties of matter or classical waves in non-Hermitian periodic potentials, i.e. in the so-called complex crystals (see, for instance, [1-31] and references therein). Among them, parity-time ($\mathcal{PT}$) symmetric complex crystals, which possess a real-valued energy spectrum below a  symmetry-breaking point \cite{Bender1,Bender2,Bender3}, have attracted a huge attention, especially since the proposal \cite{C12} and experimental realizations \cite{O1,O2,O3,O4} in optics of synthetic periodic optical media with tailored optical gain and loss regions. 
Complex crystals show rather unique scattering and transport properties as compared to ordinary (Hermitian) crystals,  such as  violation of the Friedel's law of Bragg
scattering \cite{C6,C7,C9,C10}, double refraction and nonreciprocal diffraction \cite{C12},
unidirectional Bloch oscillations \cite{C13}, mobility transition and hyper ballistic transport \cite{C30,O4}, and unidirectional invisibility \cite{C23,C24,C28,C31,O2,O3}. One among the most investigated $ \mathcal{PT}$-symmetric complex crystals is the one described by the sinusoidal complex potential $V(x)=V_0 \exp(-2ik_0x)$, which is amenable for an exact analytical study \cite{C2,C3,C6,C7,C12,C19,C20,C21,C23,C24,C25,C28,C31}. In the infinitely-extended crystal, this periodic potential is gapless and shows a countable set of spectral singularities, which are the signature of the $\mathcal{PT}$ symmetry breaking transition. For a finite crystal of length $L$ containing a finite number $N$ of unit cells, it was shown by simple coupled-mode theory that in the limit of a shallow potential the crystal is unidirectionally invisible, i.e. it is transparent and does not reflect waves when probed in one propagation direction \cite{C23,Azana}. One-way invisible crystals with
sophisticated shape and structure have been also synthesized by application of supersymmetric (SUSY) transformations
of the sinusoidal complex crystal at its symmetry
breaking point \cite{C31}. Unidirectional invisibility holds for thin and shallow enough crystals, a condition which is typically satisfied in optical experiments \cite{O2,O3}. However, for thick crystals the transparency and unidirectional reflectionless properties of the sinusoidal complex potential break down and the scattering scenario comprises three distinct regimes, as shown in Refs.\cite{C24,C28} by an exact analysis of the scattering problem involving modified Bessel functions beyond coupled-mode theory. An open question remains whether there exist $\mathcal{PT}$-symmetric complex crystals that remain unidirectionally invisible as $L \rightarrow \infty$. \par 
In this work we introduce a new class of exactly-solvable complex periodic potentials which are transparent and unidirectional reflectionless (at any degree of accuracy) over a broad frequency interval around the Bragg frequency and that remain unidirectionally invisible as the crystal length $L$ becomes infinite. Such complex crystals are super symmetrically associated to an Hermitian square potential well of length $L$ and height $\epsilon$. In the limit of small $\epsilon$, the partner complex crystal is almost unidirectional invisible, even in the $L \rightarrow \infty$ limit, and its shape differs from the complex sinusoidal potential at the symmetry breaking point, previously considered in Refs.\cite{C23,C24,C28}, mainly for a bias of the real part of the potential, which avoids breakdown of transparency as $L \rightarrow \infty$.

\section{Synthesis of the $\mathcal{PT}$-symmetric complex crystal}
Let us consider the stationary Schr\"{o}dinger equation for a quantum particle in a locally periodic and complex potential $V(x)$, which in dimensionless form reads
\begin{equation}
\hat{H} \psi \equiv -\frac{d^2 \psi}{dx^2}+V(x) \psi=E \psi
\end{equation}
where $E$ is the energy of the incident particle and $V(x)$ is the complex scattering potential with period $\Lambda$, which is nonvanishing in the interval $0<x<L$. In dimensionless units used here $\Lambda$ is taken to be of order $ \sim 1$, for example $\Lambda= \pi$. The crystal length $L$ is assumed to be an integer multiple of the lattice period $\Lambda$, i.e. $L=N \Lambda$, where $N$ is the number of unit cells in the crystal. As discussed in Refs.\cite{C24,C31},  Eq.(1) can also describe Bragg scattering of optical waves from a complex grating of period $\Lambda$ and length $L$ at frequencies close to the Bragg frequency. 
Our aim is to synthesize a complex periodic potential $V(x)$ which is almost unidirectional invisible over a broad frequency range around the Bragg frequency and that remains unidirectionally invisible in the $N \rightarrow \infty$ limit. To this aim, we use SUSY transformations (see e.g.  \cite{susy1,susy2}) to realize isospectral partner potentials, one of which being almost bidirectionally invisible and corresponding to a shallow square potential well. Let us indicate by $\hat{H}_1=-\partial_x^2+V_1(x)$  the Hamiltonian corresponding to the potential $V_1(x)$, and let $\phi(x)$ be a solution (not necessarily normalizable) to the equation $\hat{H}_{1} \phi=E_1 \phi$. The Hamiltonian $\hat{H}_1$ can be then factorized as $\hat{H}_1=\hat{B}\hat{A}+E_1$, where $\hat{A}=-\partial_x+W(x)$,  $\hat{B}=\partial_x+W(x)$, and 
\begin{equation}
W(x)=\frac{(d \phi / dx)}{\phi(x)}
\end{equation}
is the so-called superpotential. The Hamiltonian $\hat{H}=\hat{A}\hat{B}+E_1$, obtained by intertwining the operators ${\hat A}$ and $\hat{B}$, is called the partner Hamiltonian of $\hat{H}_1$. The following properties then hold:\\
(i) The potential $V(x)$ of the partner Hamiltonian $\hat{H}$ is given by
\begin{equation}
V(x)=W^2(x)-\frac{dW}{dx}+E_1=-V_1(x)+2E_1+2W^2(x)
\end{equation}
(ii) If $\hat{H}_1 \psi=E \psi$ with $E \neq E_1$, then $\hat{H} \xi=E \xi$ with
\begin{equation}
\xi(x)=\hat{A} \psi(x)=-\frac{d \psi}{dx}+W(x) \psi(x).
\end{equation}
(iii) The two linearly-independent solutions to the equation $\hat{H} \xi= E_1 \xi$ are given by
\begin{equation}
\xi_1(x)=\frac{1}{\phi(x)} \; , \; \xi_2(x)= \frac{1}{\phi(x)} \int_0^x dt \phi^2(t).
\end{equation}
Let us apply the SUSY transformation by assuming for $V_1(x)$ a shallow square potential well of width $L=N \Lambda$ and depth $\epsilon$, i.e. 
\begin{equation}
V_1(x)= \left\{
\begin{array}{cc}
0 & x<0, \; \; x>L \\
-\epsilon & 0<x<L
\end{array}
\right.
\end{equation}
Let us indicate by $k_0$ the Bragg wave number, $k_0=\pi / \Lambda$, and let us assume $E_1=k_0^2- \epsilon>0$. A solution  $\phi(x)$ to the equation $\hat{H}_1 \phi=E_1 \phi$ is given by
\begin{equation}
\phi(x)= \left\{
\begin{array}{cc}
\exp(ik_1x) & x<0 \\
\mu \cos(k_0 x-i \rho) & 0<x<L \\
\exp[ik_1(x-L)-iN \pi] & x>L
\end{array}
\right.
\end{equation}
\begin{figure}[htb]
\centerline{\includegraphics[width=12cm]{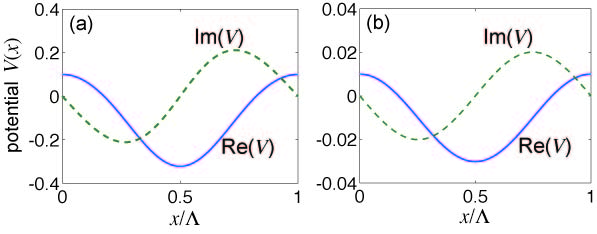}} \caption{
Behavior of the complex periodic potential $V(x)$ (real and imaginary parts), defined by Eq.(10), in the unit cell $0<x<\Lambda$ for $\Lambda=\pi$ and for (a) $\epsilon=0.1$, and (b) $\epsilon=0.01$.}
\end{figure}
where we have set $k_1= \sqrt{E_1}=\sqrt{k_0^2-\epsilon}$ and where $\rho$, $\mu$ are two real parameters that need to be determined by imposing the continuity of $\phi(x)$ and of its first derivative at $x=0$ and at $x=L=N \pi /k_0$. This yields
\begin{equation}
\rho={\rm atanh} \left( \sqrt{ 1-\frac{\epsilon}{k_0^2}} \right) \; , \;\; \mu= \frac{\sqrt{\epsilon}}{k_0}.
\end{equation}
From Eqs.(2) and (7), it follows that the superpotential $W(x)$ is given by
\begin{equation}
W(x)= \left\{
\begin{array}{cc}
i k_1 & x<0 \; , \; \; x>L \\
-k_0 \tan (k_0x-i \rho) & 0<x<L
\end{array}
\right.
\end{equation}
The potential $V(x)$ of the partner Hamiltonian is readily obtained from Eqs.(3) and (9), and reads explicitly
\begin{equation}
V(x)= \left\{ 
\begin{array}{cc}
0 & x<0 \; , \; \; x>L \\
\frac{4k_0^2}{1+\cos(2 k_0 x-2i \rho)} -\epsilon & 0<x<L
\end{array}
\right.
\end{equation}
Note that in the interval $(0,L)$ the potential $V(x)$ is locally periodic with period $\Lambda= \pi/k_0$, i.e. $k_0$ is the Bragg wave number. The real and imaginary parts of the potential, $V(x)=V_R(x)+iV_I(x)$, are given by
\begin{eqnarray}
V_R(x) & = & \frac{4 k_0^2 \left[ 1+ \cos( 2 k_0x) \cosh(2 \rho) \right]}{\left[\cos( 2 k_0x) + \cosh(2 \rho) \right]^2}-\epsilon \\
V_I(x) & = & - \frac{4 k_0^2 \sin( 2 k_0x) \sinh(2 \rho)}{\left[\cos( 2 k_0x) + \cosh(2 \rho) \right]^2}.
\end{eqnarray}
Note that, since $V_R(x)$ and $V_I(x)$ have opposite parity in the unit cell, the crystal is $\mathcal{PT}$ symmetric. A typical behavior of the real and imaginary parts of the potential is shown in Fig.1. Interestingly, in the limit $ \epsilon \rightarrow 0$ (i.e. $\rho \rightarrow \infty$), the potential $V(x)$ in the interval $0<x<L$ reduces to
\begin{equation}
V(x) \simeq 2 \epsilon \exp(-2i k_0x)-\epsilon,
\end{equation}
i.e. $V(x)$ basically coincides with the complex sinusoidal potential at the symmetry breaking point \cite{C23,C24,C28}, but with the additional bias $-\epsilon$. As discussed in the next section, such a bias prevents breakdown of the transparency of the crystal in the $L \rightarrow \infty $ limit.

\section{Scattering states, spectral reflection/transmission coefficients and unidirectional invisibility}
\subsection{Wave scattering from the square well potential $V_1(x)$}
Let us first consider the scattering properties of the square well potential $V_1(x)$, defined by Eq.(6).  This is a very simple and exactly solvable problem. For a plane wave with momentum $p$ incident from the left side of the well, the solution to the Schr\"{o}dinger equation $\hat{H}_1 \psi_p=E \psi_p$ ($E=p^2$) in the $x<0$ and $x>L$ regions is given by 
\begin{equation}
\psi_p(x)= \left\{ 
\begin{array}{c}
 \exp(ipx)+r_{1}^{(l)}(p) \exp(-ipx) \; \; x \leq 0 \\
t _{1}(p) \exp(ipx) \; \; x \geq L \\
\end{array}
\right.
\end{equation}
\begin{figure}[htb]
\centerline{\includegraphics[width=12cm]{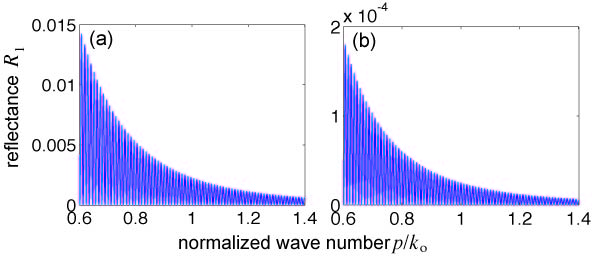}} \caption{
Behavior of the spectral reflectance $R_1(p)$ of the square potential well [Eq.(6)] for $N=100$, $\Lambda=\pi$ (i.e. $k_0=1$) and for (a) $\epsilon=0.1$,  (b) $\epsilon=0.01$}
\end{figure}
where $t_1(p)$ and $r_1^{(l)}(p)$ are the transmission and reflection (for left incidence) coefficients, respectively. Similarly, for a plane wave with momentum $p$ incident from the right side of the well, the solution to the Schr\"{o}dinger equation $\hat{H}_1 \psi_p=E \psi_p$ ($E=p^2$) in the $x<0$ and $x>L$ regions is given by 
\begin{equation}
\psi_p(x)= \left\{ 
\begin{array}{c}
 \exp(-ipx)+r_{1}^{(r)}(p) \exp(ipx) \; \; x \geq L \\
t _{1}(p) \exp(-ipx) \; \; x \leq 0 \\
\end{array}
\right.
\end{equation}
where  $r_1^{(r)}(p)$ is  the reflection coefficient for right incidence. In the well region $0<x<L$ the solution $\psi_p(x)$ is given by a superposition of plane waves $\exp(iqx)$ and $\exp(-iqx)$, where we have set
\begin{equation}
q= \sqrt{p^2+\epsilon}.
\end{equation}
The amplitudes of plane waves, as well as the expressions of the spectral transmission [$t_1(p)$] and reflection [$r_1^{(l,r)}(p)$] coefficients, are readily obtained by imposing the continuity of $\psi_p(x)$ and of its first derivative at $x=0$ and $x=L$. This yields
\begin{eqnarray}
t_1(p) & = & \frac{2pq \exp(-ipL)}{2pq \cos(qL)-i(p^2+q^2) \sin (qL)}  \\
r_{1}^{(l)}(p) & = & \frac{i \epsilon \sin(qL)}{2pq \cos(qL)-i(p^2+q^2) \sin (qL)}\\
r_{1}^{(r)}(p) & = & r_{1}^{(l)}(p) \exp(-2ipL)
\end{eqnarray}
The reflectance is equal for both left and right incidence sides and it is given by
\begin{equation}
R_1(p)=|r_{1}^{(l,r)}(p)|^2=\frac{\epsilon^2 \sin^2(qL)}{-\epsilon^2 \cos^2(qL)+(2p^2+\epsilon)^2}
\end{equation}
\begin{figure}[htb]
\centerline{\includegraphics[width=14cm]{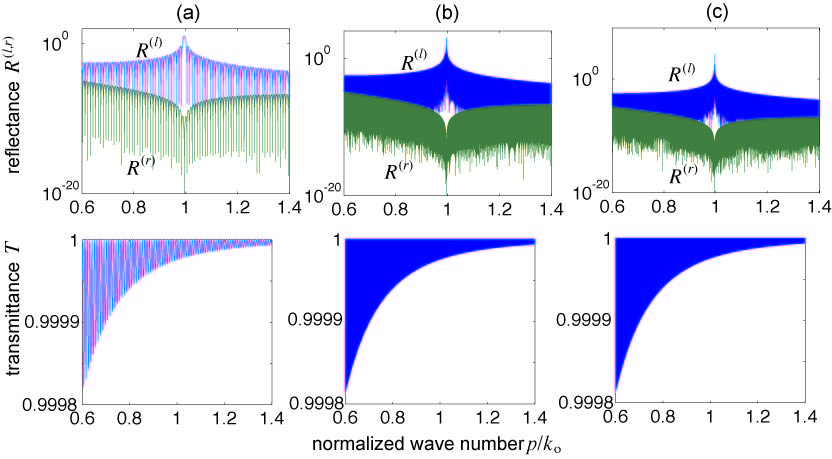}} \caption{
Behavior of the spectral reflectances $R^{(l,r)}(p)$ (for left and right incidence sides, upper plots) and transmittance $T(p)$ (lower plots) of the complex crystal $V(x)$, defined by Eq.(10), for  $\Lambda=\pi$, $\epsilon=0.01$ and for increasing number $N$ of cells: (a) $N=100$, (b) $N=1000$, and (c) $N=5000$. Note that the spectral reflectances are plotted on a logarithmic scale.}
\end{figure}
whereas the transmittance is given by $T_1(p)=1-R_1(p)$. Note that $R_1(p) \leq \epsilon^2/[4p^2(p^2+\epsilon)]$, so that for a fixed value of $p_0>0$ one has $R_1(p) \rightarrow 0$, $T_1(p) \rightarrow 1$  as $\epsilon \rightarrow 0$ uniformly in the interval $p  \in (p_0, \infty)$, regardless of the value of $L$ \footnote{More precisely, for a fixed value of $p_0>0$ and for a given (arbitrarily small) value $\eta>0$, one can find a nonvanishing value of $\epsilon$ (independent of $L$) such 
that $R_1(p)< \eta$ for any $L$ and for any $p$ in the range $p \in (p_0, \infty)$.
}. This means that, for a sufficiently small value of $\epsilon$, the potential well is almost reflectionless from both sides of incidence in a wide interval of wave numbers $p$ around the Bragg wave number $p=k_0$. As an example, Fig.2 shows typical behaviors of the reflectance $R_1(p)$ as $p$ varies in the range $(0.6 k_0,1.4k_0)$ for two values of $\epsilon$. 
\subsection{Wave scattering from the $\mathcal{PT}$-symmetric complex potential $V(x)$}
The scattering properties of the complex crystal defined by Eq.(10) are readily obtained from those of the partner square well potential $V_1(x)$ using the property (ii) of SUSY stated in the previous section.  For a plane wave with momentum $p$ incident from the left side of the crystal, the solution to the Schr\"{o}dinger equation $\hat{H} \xi_p=E \xi_p$ ($E=p^2$) in the $x<0$ and $x>L$ regions is given by 
\begin{equation}
\xi_p(x)= \alpha \left\{ 
\begin{array}{c}
 \exp(ipx)+r^{(l)}(p) \exp(-ipx) \; \; x \leq 0 \\
t (p) \exp(ipx) \; \; x \geq L \\
\end{array}
\right.
\end{equation}
where $t(p)$ and $r^{(l)}(p)$ are the transmission and reflection (for left incidence) coefficients, respectively, and $\alpha$ is an arbitrary non-vanishing constant. On the other hand, according to the property (ii) stated in the previous section for $E \neq E_1$, i.e. for $p \neq k_1=\sqrt{k_0^2-\epsilon}$, one has $\xi_p(x)=-(d \psi_p/dx)+W(x) \psi_p(x)$. From Eqs.(14) and (21), taking into account that $W(x)=ik_1$ for $x<0$ and $x>L$, it follows that $\alpha=i(k_1-p)$ and
\begin{equation}
t(p)=t_1(p) \; , \; \; r^{(l)}(p)=r_{1}^{(l)}(p) \frac{k_1+p}{k_1-p}
\end{equation}
Similarly, from the problem of a plane wave with momentum $p$ incident from the right side of the crystal one obtains
\begin{equation}
t(p)=t_1(p) \; , \; \; r^{(r)}(p)=r_{1}^{(r)}(p) \frac{k_1-p}{k_1+p}.
\end{equation}
The case $E=E_1$, i.e. $p=k_1$, can be analyzed by considering the limit of Eqs.(22) and (23) as $p \rightarrow k_1$. In particular, from Eq.(23) one has $r^{(r)}(p) \rightarrow 0$ as $p \rightarrow k_1$, whereas from Eqs.(18) and (22) a 0/0 limit is obtained, which yields after some calculations  
\begin{equation}
r^{(l)}(p) \rightarrow   -i \epsilon L \frac{k_1}{k_0}
\end{equation}
as $p \rightarrow k_1$. From the above results, the following scattering properties of the complex crystal, with potential $V(x)$ given by Eq.(10), can be stated:\\
(i) The transmission coefficient $t(p)$ of the complex crystal is the same as that $t_1(p)$ of a square well of width $L$ and height $\epsilon$, see Eq.(17).\\
(ii) The spectral reflectances $R^{(l)}(p)=|r^{(l)}(p)|^2$ and $R^{(r)}(p)=|r^{(r)}(p)|^2$ for left and right hand incidence sides are related to the reflectance $R_1(p)$ of the square well [Eq.(20)] by the simple relations
\begin{equation}
R^{(l)}(p)=R_1(p) \left( \frac{k_1+p}{k_1-p}\right)^2 \; , \; \; R^{(r)}(p)=R_1(p) \left( \frac{k_1-p}{k_1+p}\right)^2.
\end{equation}
Note that the spectral transmittance $T(p)=|t(p)|^2$ and reflectances $R^{(l)}(p)$, $R^{(r)}(p)$ satisfy the generalized unitarity relation 
\begin{equation}
T-1=\sqrt{R^{(l)}R^{(r)}},
\end{equation}
 according to general results on scattering in $\mathcal{PT}$-symmetric potentials \cite{C31,uff0,uff1,uff2}.\\
 (iii) For a fixed momentum $p_0>0$ and for a given small parameter $\eta>0$, one can find a non-vanishing value $\epsilon>0$, independent of $L$, such that $R^{(r)}(p)< \eta$ and $|T(p)-1|< \eta$ uniformly in the interval $(p_0, \infty)$, regardless of the value of $L$. Moreover, the reflectance for left side incidence shows a peak at $p=k_1=\sqrt{k_0^2-\epsilon}$, which secularly grows with the crystal thickness $L$ like $\sim L^2$, namely one has
 \begin{equation}
 R^{(l)}(p=k_1)=\epsilon^2 L^2 \left(  1-\frac{\epsilon} {k_0^2} \right)
 \end{equation}
 \begin{figure}[htb]
\centerline{\includegraphics[width=14cm]{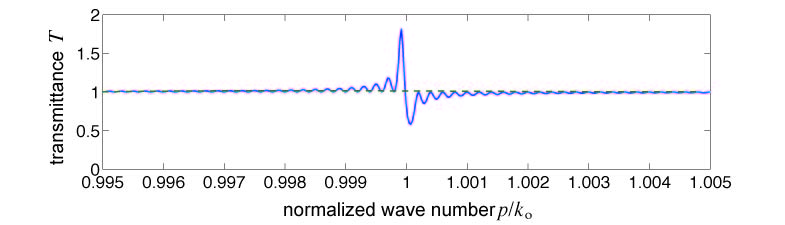}} \caption{
Numerically-computed behavior of the spectral transmittance $T(p)$  of the complex sinusoidal potential $V(x)=2 \epsilon \exp(-2ik_0x)$ (solid curve), and of the shifted
complex sinusoidal potential$V(x)=2 \epsilon \exp(-2ik_0x)-\epsilon$ (dashed curve) for  $\Lambda=\pi$, $\epsilon=0.01$ and $N=5000$.}
\end{figure}
 This means that, for a sufficiently small value of $\epsilon$, the potential (10) is almost transparent and unidirectional reflectionless in a broad range of wave number $p$ around the Bragg wave number $k_0$ and for an {\it arbitrary} crystal thickness $L$. This is shown in Fig.3, where typical behaviors of spectral transmittance $T(p)$ and reflectances $R^{(l,r)}(p)$ for increasing values of $L=N \Lambda$ are depicted. Note that, as $N$ increases the spectral reflectance $R^{(r)}(p)$ remains smaller than $ \sim 2 \times 10^{-4}$, the transmittance $T(p)$ remains close to 1, whereas the spectral reflectance $R^{(l)}(p)$ shows a peak at $p=k_1 \simeq k_0$, which increases as $N$ is increased [according to Eq.(27)]. The scattering properties of the potential (10) are thus very distinct than the ones of the complex sinusoidal potential at the symmetry breaking point,
 $V(x)=2 \epsilon \exp(-2ik_0x)$, for which transparency is lost as $N$  is increased \cite{C24}. This is shown in Fig.4, where the solid line shows the behavior of spectral transmittance of the complex sinusoidal potential for $\epsilon=0.01$, $\Lambda=\pi$ and $N=5000$, i.e. for the same parameter values as in Fig.3(c). Note that near the Bragg wave number $p \simeq k_0$ the transmittance $T$ greatly deviates from 1. As discussed in Ref.\cite{C24}, such a deviation is the signature of the spectral singularity of the complex sinusoidal potential that arises in the $L \rightarrow \infty$ limit. For a small value of $\epsilon$, the potential (10) is well approximated by the {\it shifted} complex sinusoidal potential $V(x) = 2 \epsilon \exp(2ik_0x)-\epsilon$, see Eq.(13). Hence transparency of the complex sinusoidal potential is expected to be restored provided that the {\it bias} $-\epsilon$ is added to the potential. This is shown in Fig.4, where the dashed curve depicts the numerically-computed behavior of the spectral transmittance for the {\it shifted} sinusoidal complex potential $V(x) = 2 \epsilon \exp(2ik_0x)-\epsilon$. As one can appreciate from the figure, the addition of the bias $-\epsilon$ to the complex sinusoidal potential restores the transparency. 
 
\section{Conclusion}
In this work Bragg scattering in a new class of super-symmetrically synthesized $\mathcal{PT}$-symmetric complex crystals has been analytically investigated. The crystal turns out to be almost transparent and unidirectionally reflectionless, i.e. one-way invisible. Unidirectional invisibility  has been predicted and recently observed in the sinusoidal complex potential at the symmetry breaking point, see  \cite{C2,C3,C23,C24,C28,C31,O3}. However, while transparency is lost for the complex sinusoidal potential for thick crystals \cite{C24}, in the super-symmetrically synthesized complex crystal considered in this work unidirectional reflectionless and transparency properties hold regardless of the crystal thickness, i.e. they persist in the limit $L \rightarrow \infty$. The reason thereof is that the spectral transmittance of the super-symmetrically synthesized complex crystal is the same as that of a shallow square well potential, which remains close to one regardless of the thickness $L$ of the well. In the limit of a very shallow potential well, we have shown that the  
super-symmetrically associated complex crystal reduces to the complex sinusoidal potential at the symmetry breaking point, {\it but} with an additional bias of the potential, which prevents breakdown of transparency in the $L \rightarrow \infty$ limit.



\section*{References}
\begin{thebibliography}{10}

\bibitem{C1}
N. Hatano and D.R. Nelson 1996 {\it Phys. Rev. Lett.} {\bf 77}  570

\bibitem{C2}
Cannata F, Junker G and Trost J 1998  {\it Phys. Lett. A}  {\bf 246}  219

\bibitem{C3}
Bender C M, Dunne G V  and  Meisinger P N 1999 {\it Phys. Lett. A} {\bf 252} 272 

\bibitem{C4}
Cervero J M 2003 {\it Phys. Lett. A} {\bf A317} 26 

\bibitem{C5}
Shin K C 2004  {\it J. Phys. A: Math. Gen.} {\bf 37} 8287 

\bibitem{C6}
 Berry M V 1998 {\it J. Phys. A} {\bf 31} 3493 
 
 \bibitem{C7}
 Berry M V and  O'Dell D H J 1998 {\it J. Phys. A} {\bf 31} 2093

\bibitem{C8}
Berry M V 2008 {\it J. Phys. A} {\bf 41}  244007

\bibitem{C9}
Oberthaler M K, Abfalterer R, Bernet S, Schmiedmayer J and Zeilinger A 1996 
{\it Phys. Rev. Lett.} {\bf 77} 4980

\bibitem{C10}
Keller C, Oberthaler M K, Abfalterer R, Bernet S,
Schmiedmayer J and Zeilinger A 1997 {\it Phys. Rev. Lett.} {\bf 79} 3327

 \bibitem{C11}
 St\"{u}tzle R,  G\"{o}bel M C, H\"{o}rner Th, Kierig E, Mourachko I, Oberthaler M K, Efremov M A,  Fedorov M V, Yakovlev V P, van Leeuwen K A H and Schleich W P 2005 {\it Phys. Rev. Lett.} {\bf 95} 110405

\bibitem{C12}
Makris K G,  El-Ganainy R, Christodoulides D N and 
Musslimani Z H 2008 {\it Phys. Rev. Lett.} {\bf 100} 103904

\bibitem{C13}
Longhi S 2009 {\it Phys. Rev. Lett.} {\bf 103} 123601 

\bibitem{C14}
Longhi S 2009 {\it Phys. Rev. B} {\bf 80} 235102 

\bibitem{C15}
Mostafazadeh A 2009 {\it Phys. Rev. Lett.} {\bf 102} 220402 

\bibitem{C16}
Jin L and Song Z 2009 {\it Phys. Rev. A} {\bf 80}  052107

\bibitem{C17}
Jin L and  Song Z 2010 {\it Phys. Rev. A} {\bf 81}  032109

\bibitem{C18}
 Ramezani H,  Kottos T, El-Ganainy R and  Christodoulides D N 2010
{\it Phys. Rev. A} {\bf 82}, 043803 

\bibitem{C19}
Makris K G, El-Ganainy R, Christodoulides D N  and Musslimani Z H 
2010 {\it Phys. Rev. A} {\bf 81} 063807 

 \bibitem{C20}
 Midya B, Roy B and  Roychoudhury R 2010 {\it Phys. Lett. A} {\bf 374} 2605
 
\bibitem{C21}
Longhi S 2010 {\it Phys. Rev. A} {\bf 81} 022102

\bibitem{C22}
Longhi S 2010  {\it Phys. Rev. Lett.} {\bf 105} 013903 

\bibitem{C23}
Lin Z,  Ramezani H,  Eichelkraut T, Kottos T, Cao H and Christodoulides D N
2011 {\it Phys. Rev. Lett.} {\bf 106} 213901 

\bibitem{C24}
Longhi S 2011 {\it J. Phys. A} {\bf 44} 485302

\bibitem{C25}
 Graefe E M and Jones H F 2011
{\it Phys. Rev. A} {\bf 84} 013818 

\bibitem{C26}
Longhi S, Cannata F  and Ventura A 2011 {\it Phys. Rev. B} {\bf 84}  235131

\bibitem{C27}
Longhi S, Della Valle G and Staliunas K 2011 {\it Phys. Rev. A} {\bf 84} 042119

\bibitem{C28}
Jones H F 2012 {\it J. Phys. A} {\bf 45} 135306

\bibitem{C29}
Miri M A,  Regensburger A,  Peschel U and Christodoulides D N 2012 {\it Phys. Rev. A} {\bf 86} 023807

\bibitem{C30}
Della Valle G and Longhi S 2013 {\it Phys. Rev. A} {\bf 87} 022119

\bibitem{C31}
Midya B 2014 {\it Phys. Rev. A} {\bf 89} 032116 

\bibitem{Bender1}
Bender C M and Boettcher S 1998 {\it Phys. Rev. Lett.} {\bf 80} 5243

\bibitem{Bender2}
Bender C M 2007 {\it Rep. Prog. Phys.} {\bf 70} 947 

\bibitem{Bender3}
Mostafazadeh A 2010 {\it Phys. Scr.} {\bf 82} 038110

\bibitem{O1} 
Feng L, Ayache M, Huang J,  Xu Y L,  Lu M H,
 Chen Y F, Fainman Y, and Scherer A 2011 {\it Science} {\bf 333} 729 

\bibitem{O2}
 A. Regensburger, C. Bersch, M.-A. Miri, G. Onishchukov,
D. N. Christodoulides, and U. Peschel, {\it Nature} {\bf 488}, 167 (2012).
 
\bibitem{O3} 
L. Feng, Y.-L. Xu, W. S. Fegadolli, M.-H. Lu, J. E. B. Oliveira,
V. R. Almeida, Y.-F. Chen, and A. Scherer, {\it Nature Mater.} {\bf 12}, 108
(2013).

\bibitem{O4}
Eichelkraut T, Heilmann R, Weimann S, St\"{u}tzer, Dreisow F, Christodoulides D N,
Nolte S  and Szameit A 2013 {\it Nature Commun.} {\bf 4} 2533

\bibitem{Azana}
Kulishov M, Laniel J M,  Belanger N, Azana J and
Plant D V 2005 {\it Opt. Express} {\bf 13} 3068 

\bibitem{susy1}
Cooper F, Khare A and Sukhatme U 1995 {\it Phys. Rep.} {\bf 251} 267

\bibitem{susy2}
Dunne G  and Feinberg J 1998 {\it Phys. Rev. D} {\bf 57} 1271


\bibitem{uff0}
Cannata F, Dedonder J P and Ventura A 2007 {\it Ann. Phys.}
{\bf 322} 397

\bibitem{uff1}
Ge L, Chong  Y D and Stone A D 2012 {\it Phys. Rev. A} {\bf 85}
023802 

\bibitem{uff2}
Ahmed Z 2013 {\it Phys. Lett. A} {\bf 377}  957


\end{thebibliography}
\end{document}